\documentclass[aps,prl,superscriptaddress,twocolumn]{revtex4}
\usepackage{amssymb}
\usepackage{graphicx}
\usepackage{amsmath}
\usepackage[colorlinks=true,linkcolor=blue,citecolor=blue,urlcolor=blue]{hyperref}

\begin{document}

\title{R\'enyi Entropy Dynamics and Lindblad Spectrum for Open Quantum System}
\author{Yi-Neng Zhou}
\thanks{They contribute equally to this work. }
\affiliation{Institute for Advanced Study, Tsinghua University, Beijing 100084, China}
\author{Liang Mao}
\thanks{They contribute equally to this work. }
\affiliation{Institute for Advanced Study, Tsinghua University, Beijing 100084, China}
\affiliation{Department of Physics, Tsinghua University, Beijing 100084, China}
\author{Hui Zhai}
\email{hzhai@tsinghua.edu.cn}
\affiliation{Institute for Advanced Study, Tsinghua University, Beijing 100084, China}
\date{\today}

\begin{abstract}
In this letter we point out that the Lindblad spectrum of a quantum many-body system displays a segment structure and exhibits two different energy scales in the strong dissipation regime. One energy scale determines the separation between different segments, being proportional to the dissipation strength, and the other energy scale determines the broadening of each segment, being inversely proportional to the dissipation strength. Ultilizing a relation between the dynamics of the second R\'enyi entropy and the Lindblad spectrum, we show that these two energy scales respectively determine the short- and the long-time dynamics of the second R\'enyi entropy starting from a generic initial state. This gives rise to opposite behaviors, that is, as the dissipation strength increases, the short-time dynamics becomes faster and the long-time dynamics becomes slower. We also interpret the quantum Zeno effect as specific initial states that only occupy the Lindblad spectrum around zero, for which only the broadening energy scale of the Lindblad spectrum matters and gives rise to suppressed dynamics with stronger dissipation. We illustrate our theory with two concrete models that can be experimentally verified.    

\end{abstract}

\maketitle

For a closed quantum system, the energy spectrums of Hamiltonian fully determine the time scales of its dynamics. For an open quantum system, when the environment is treated by the Markovian approximation, the couplings between system and environment are controlled by a set of dissipation operators. In this case, the dynamics of the system is governed by the Lindblad equation which contains the contributions from both the Hamiltonian and the dissipation operators \cite{open}. Obviously, the spectrum of the Hamiltonian alone can no longer determine the time scales of the entire dynamics, and a natural question is then what energy scales set the time scales of dynamics of an open quantum system. 

There are various directions to approach this issue, and the answer also relies on what type of dynamics that we are concerned with. Here let us focus on the dissipation driven dynamics. There are still different physical intuitions from different perspectives. One intuition is from the perturbation theory when the dissipation strength is weaker compared with the typical energy scales of the Hamiltonian \cite{Pan}. In this regime, by treating the dissipation perturbatively, it leads to a scenario that the dissipation dynamics becomes faster when the dissipation strength is stronger. Another intuition is from the studies of the quantum Zeno effect \cite{Zeno_paradox,Zeno1990,quantum_Zeno,Zeno_review}, which states that frequent measurements can slow down the dynamics, provided that the typical time interval between two successive measurements are shorter than the intrinsic time scale of the system. Since the measurement can also be understood in term of dissipations in the Lindblad master equation, it provides another scenario that the dissipation dynamics is suppressed when the dissipation becomes stronger, in the regime that the dissipation strength is stronger compared with the typical energy scales of the Hamiltonian. It seems that these two scenarios respectively work on different parameter regimes and the results are also opposite to each other. It will be interesting to see that there actually exists a framework that can unify these two scenarios. 

When a system is coupled to a Markovian environment, the entropy of the system will increase in time. The entropy dynamics of an open quantum many-body system is a subject that attracts lots of interests recently \cite{entang_review,Qi_Cricuit,Chen,Kitaev,YuChen,Zhai}. In this letter, we address the issue of typical time scales of the entropy increasing dynamics of a quantum many-body system coupled to a Markovian invironment, and especially, we should focus on the second R\'enyi entropy, for the reason that will be clear below, and answer the question whether the entropy dynamics is faster or slower when the dissipation strength increases. 

\begin{figure}[t] 
    \centering
    \includegraphics[width=0.48\textwidth]{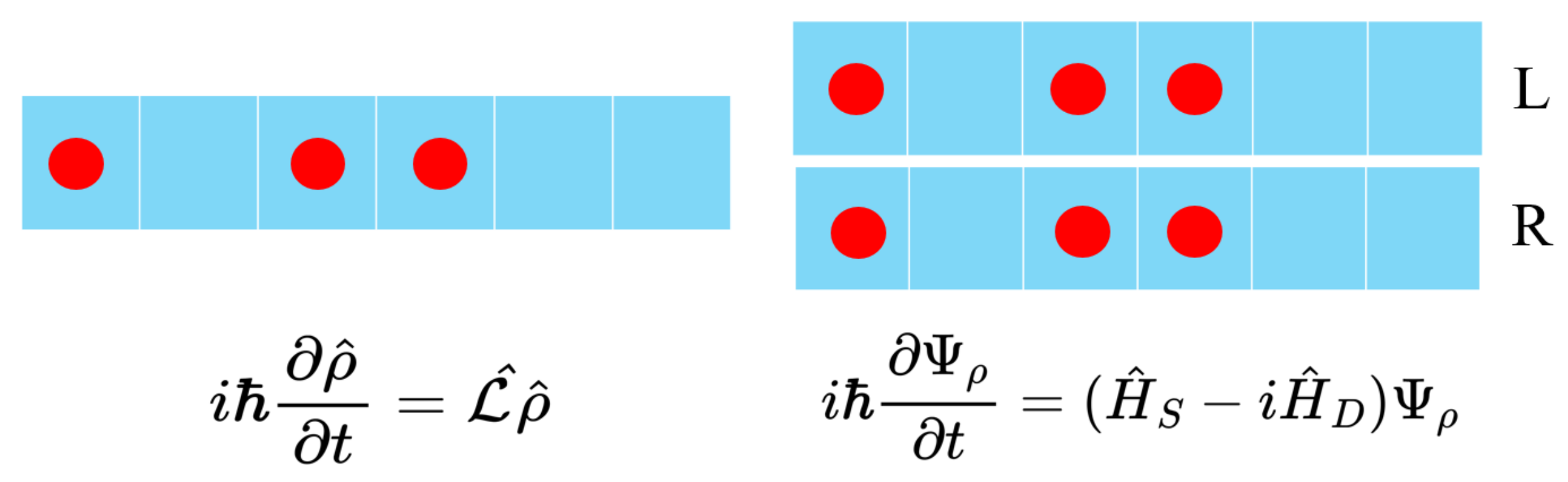}
    \caption{Schematic of the mapping between the Lindblad equation (left) and the Sch\"odinger like equation in a doubled system (right). Here $\hat{L}\hat{\rho}$ denotes the r.h.s. of Eq. \ref{Lindblad}. }
     \label{schematic}
\end{figure} 

Our studies are based on a mapping between the Lindblad master equation and a non-unitary evolution of wave function in a doubled space, as shown in Fig. \ref{schematic}. Let us first review this mapping \cite{Operator_Schmidt,Mixed_state}. Considering a density matrix $\hat{\rho}$, and given a set of complete bases $\{|n\rangle\}, (n=1,\dots,\mathcal{D}_\text{H})$ of the Hilbert space with dimension $\mathcal{D}_\text{H}$ (say, the eigenstates of the Hamiltonian $\hat{H}$ with eigenenergies $E_n$), the density matrix $\hat{\rho}$ can be expressed as $\hat{\rho}=\sum_{mn}\rho_{mn}|m\rangle\langle n|$. By the operator-to-state mapping, we can construct a wave function $\Psi_\rho=\sum_{mn}\rho_{mn}|m\rangle \otimes |n\rangle$, which contains exact the same amount information as $\hat{\rho}$. Here $\Psi_\rho$ is a wave function on a system whose size is doubled compared to the original system, and we will refer these two copies of original system as the ``left" (L) and the ``right" (R) systems. Under this mapping, for instance, a density matrix of a pure state $\hat{\rho}=|\psi\rangle\langle \psi |$ is mapped to a product state $\Psi_\rho=|\psi\rangle\otimes|\psi\rangle$ in the double system, and a thermal density matrix at temperature $T$ as $\hat{\rho}=\sum_n e^{- E_n/(k_\text{b}T)}|n\rangle\langle n|$ is mapped to a thermofield double state at temperature $T/2$ as $\Psi_\rho=\sum e^{- E_n/(k_\text{b}T)}|n\rangle\otimes|n\rangle$ in the double system. 

For an open system coupled to a Markovian environment, the density matrix obeys the Lindblad master equation given by        
\begin{equation}
\hbar\frac{d\hat{\rho}}{dt}=-i[\hat{H},\hat{\rho}]+\sum\limits_{\mu}\gamma_\mu\left(2\hat{L}_\mu\hat{\rho}\hat{L}_\mu^\dag-\{\hat{L}^\dag_\mu\hat{L}_\mu,\hat{\rho}\}\right), \label{Lindblad}
\end{equation}
where $\hat{L}_\mu$ stand for a set of dissipation operators, and $\gamma_\mu$ are their corresponding dissipation strengths. After the mapping, the wave function $\Psi_\rho$ in the double system satisfies a Schr\"odinger-like equation 
\begin{equation}
i\hbar\frac{d\Psi_\rho}{dt}=\left(\hat{H}_\text{s}-i\hat{H}_\text{d}\right)\Psi_\rho. 
\end{equation}
Here $\hat{H}_\text{s}$ is the Hermitian part of the Hamiltonian determined by system itself, and it is given by 
\begin{equation}
\hat{H}_\text{s}=\hat{H}_\text{L}\otimes\hat{I}_\text{R}-\hat{I}_\text{L}\otimes\hat{H}^\text{T}_\text{R},
\end{equation}
where operators with subscript ``L" and ``R" respectively stand for operators acting on the left and the right systems, and ``T" stands for the transpose, and $\hat{I}$ represents the identity operator. $-i\hat{H}_\text{d}$ is the non-Hermitian part of the Hamiltonian determined by the dissipation operators, which is given by
\begin{align}
\hat{H}_\text{d}=\sum_\mu\gamma_\mu&\left[-2\hat{L}_{\mu,\text{L}}\otimes\hat{L}^\text{*}_{\mu,\text{R}}\right.\nonumber\\ &\left. +(\hat{L}^\dag_{\mu}\hat{L}_{\mu})_\text{L}\otimes\hat{I}_\text{R}+\hat{I}_{\text{L}}\otimes(\hat{L}^\dag_{\mu}\hat{L}_{\mu})^\text{*}_\text{R}\right],
\end{align}
where the superscript $\text{*}$ stands for taking complex conjugation. 
We can diagnolize this non-Hermitian Hamiltonian $\hat{H}_\text{s}-i\hat{H}_\text{d}$, which leads to a set of eigenstates as
\begin{equation}
(\hat{H}_\text{s}-i\hat{H}_\text{d})|\Psi^l_\rho\rangle=\epsilon_l|\Psi^l_\rho\rangle,
\end{equation}
where $\epsilon_l$ is in general a complex number, and we denote them as $\epsilon_l=\alpha_l-i\beta_l$. This spectrum, originated from the Lindblad equation, is referred to as the Lindblad spectrum. The full Lindblad spectrum has been studied for a number of models before \cite{Prosen1,Prosen2,Universal_spectra,tenfold,local_random_Liouvillians,Wang}. Here we would like to make several useful comments on the Lindblad spectrum. i) $\alpha_l$ and $-\alpha_l$ always appear in pairs in the spectrum; ii) $\beta_l$ is always non-negative; iii) If $\hat{L}_\mu$ are all hermitian, there always exists a zero-energy eigenstate with $\epsilon_l=0$, and this eigenstate is labelled as $l=0$ and is given by $|\Psi^{l=0}_\rho\rangle=\frac{1}{\sqrt{\mathcal{D}_\text{H}}}\sum_{n}|n\rangle\otimes|n\rangle$. 

\begin{figure}[t] 
    \centering
    \includegraphics[width=0.48\textwidth]{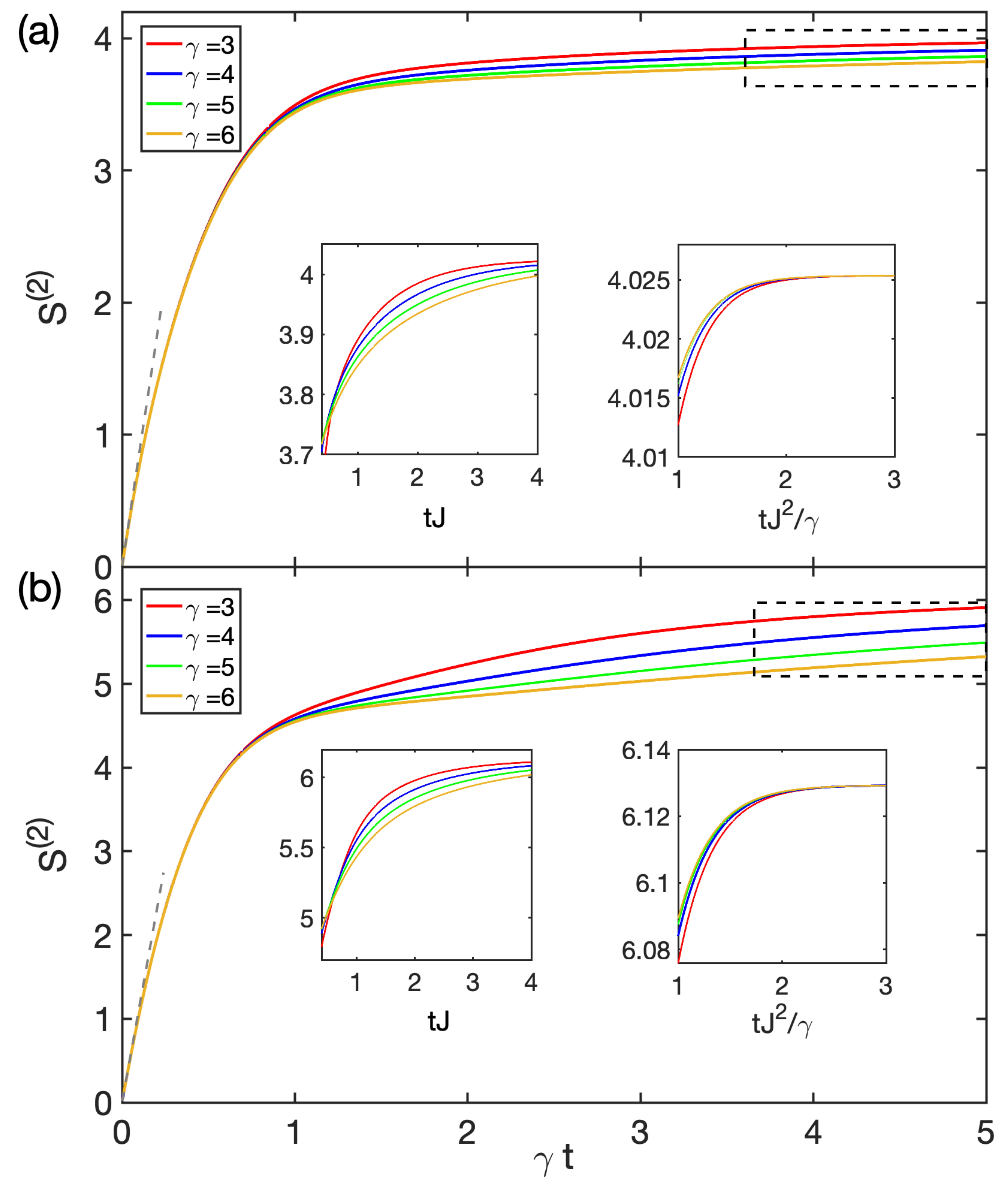}
    \caption{The dynamics of the second R\'enyi entropy $S^{(2)}$ as a function of $t\gamma$. $\gamma$ is the dissipation strength. Different curves have different $\gamma$ in unit of $J$. The inset show the long-time behavior of $S^{(2)}$ as functions of $tJ$ and $tJ^2/\gamma$. The dashed line is a fitting of initial slop based on Eq. \ref{short-time1}. (a) is for the Bose-Hubbard model with $U=J$ and the number of sites $L=6$, and the number of bosons $N=3$. (b) is for hard core bosons model with $V=J$, $L=8$ and $N=4$. The initial state is taken as the ground state of $\hat{H}$.  }
     \label{dynamics_general}
\end{figure}

\textit{R\'enyi Entropy and Lindblad Spectrum.} Here we bring out a close relation between the dynamics of the second R\'enyi entropy and the Lindblad spectrum. For any density matrix $\hat{\rho}(t)$, the second R\'enyi entropy $S^{(2)}(t)$ is given by
\begin{equation}
e^{-S^{(2)}}=\text{Tr}(\hat{\rho}^2)=\sum\limits_{mn}\rho_{mn}(t)\rho_{nm}(t). \label{Renyi}
\end{equation}
On the other hand, in the double system, the total amplitude of the wave function is given by 
\begin{equation}
|\Psi_\rho|^2=\sum\limits_{mn}\rho_{mn}(t)\rho^*_{mn}(t). \label{wf}
\end{equation}
Since the density matrix is always Hermitian, it gives $\rho_{nm}(t)=\rho^*_{mn}(t)$, and therefore, we have
\begin{equation}
e^{-S^{(2)}}=|\Psi_\rho|^2. \label{entropy-amplitude}
\end{equation}
An initial state $\Psi_\rho(0)$ in the double space can be expanded as $\Psi_\rho(0)=\sum_{l}c_l |\Psi^l_\rho\rangle$, the subsequent evolution is given by 
\begin{equation}
\Psi(t)=e^{-i\hat{H}_\text{s}t-\hat{H}_\text{d}t}|\Psi_\rho(0)\rangle=\sum_l c_l e^{-i\alpha_l t-\beta_l t}|\Psi^n_\rho\rangle \label{wf_time_dependence}
\end{equation}
and therefore
\begin{equation}
e^{-S^{(2)}}=|\Psi_\rho|^2=\sum\limits_{n}|c_l|^2 e^{-2\beta_l t}. 
\end{equation}
Since the evolution in double system is non-unitary and all $\beta_l$ are non-negative, the total amplitude of the wave function always decays in time. Hence, by this entropy-amplitude relation Eq. \ref{entropy-amplitude}, the decaying of $|\Psi_\rho|^2$ gives rise to the increasing of $S^{(2)}$. Note that for any initial density matrix with trace unity and for hermitian $\hat{L}_\mu$, $c_{l=0}$ always equals $1/\sqrt{\mathcal{D}_\text{H}}$. This mode always does not decay in time because $\beta_{l=0}=0$. If there is no other eigenmodes with $\beta_l=0$, $l=0$ mode is the only remaining mode at infinite long time, which gives a maximum second R\'enyi entropy $\log \mathcal{D}_\text{H}$. Before reaching that limit, the imaginary parts of the Lindblad spectrum of occupied states determine the time scales of the R\'enyi entropy dynamics. Our discussion below will be based on this connection.

\textit{Models.} Although our discussion below is quite general for quantum many-body systems, we illustrate the results with two concrete models. The first model is the Bose-Hubbard model, which reads
\begin{equation}
\hat{H}=-J\sum\limits_{\langle ij\rangle}(\hat{b}^\dag_i\hat{b}_j+\text{h.c.})+\frac{U}{2}\sum\limits_{i}\hat{n}_i(\hat{n}_i-1),
\end{equation} 
where $\hat{b}_i$ is the boson annihilation operator at site-$i$, and $\hat{n}_i=\hat{b}^\dag_i\hat{b}_i$ is the boson number operator at site-$i$. $\langle ij\rangle$ denotes nearest neighbor sites. $J$ and $U$ are respectively the hopping and the on-site interaction strengths. For the second model, we consider hard-core bosons, which prevent two bosons to occupy the same site. In addition, we introduce the nearest-neighbor repulsion, and the model reads
\begin{equation}
\hat{H}=-J\sum\limits_{\langle ij\rangle}(\hat{b}^\dag_i\hat{b}_j+\text{h.c.})+V\sum\limits_{\langle ij\rangle}\hat{n}_i\hat{n}_j.
\end{equation} 
In one-dimension, these two models are quite different, because the second model can be mapped to a spinless fermion model with nearest neighbor repulsion, and can also be mapped to a spin model with nearest neighbor couplings, but the first model cannot. In both cases, we take all $\hat{n}_i$ as the dissipation operators and we set the dissipation strengthes uniformly as $\gamma$. In the numerical results shown below, we have choose $J\sim U$ or $J\sim V$ such that $J$ sets the typical energy scale of the Hamiltonian part, and therefore, strong and weak dissipations respectively mean $\gamma/J> 1$ or $\gamma/J< 1$.
Below we will show that both models exhibit similar features, which supports that our results are quite universal.  

\begin{figure}[t] 
    \centering
    \includegraphics[width=0.48\textwidth]{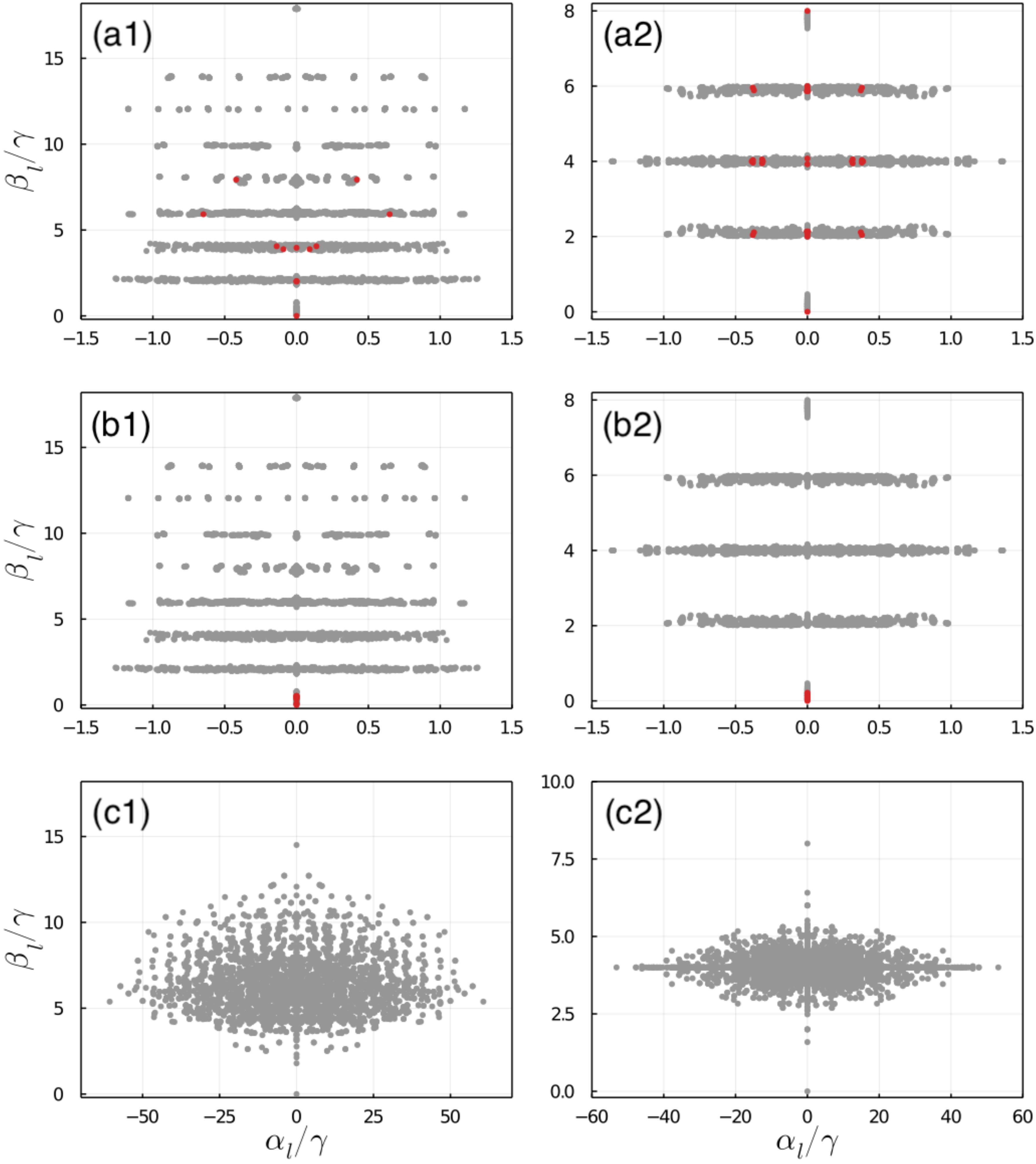}
    \caption{The Lindblad spectrum for strong dissipation case (a1,a2,b1,b2) with $\gamma=5J$ and for weak dissipation case (c1,c2) with $\gamma=0.2J$. The red points mark the eigenstates with significant occupation ($|c_l|^2\geqslant 1/\mathcal{D}_\text{H}$) by the initial state. For (a1) and (a2) in the first raw, the initial state is taken as $\Psi_\rho=|\psi_\text{g}\rangle\otimes|\psi_\text{g}\rangle$, where $|\psi_\text{g}\rangle$ is the ground state of $\hat{H}$. For (b1) and (b2) in the second raw, the initial states are taken as the zero-energy eigenstate of $\hat{H}_\text{d}$, that are $|111000\rangle$ for (b1) and $|11110000\rangle$ for (b2) in Fock bases.  The left column (a1,b1,c1) are for the Bose-Hubbard model with $U=J$ and the number of sites $L=6$, and the number of bosons $N=3$. The right column (a2,b2,c2) are for hard core bosons model with $V=J$, $L=8$ and $N=4$. }
     \label{spectrum}
\end{figure}

\textit{Dynamics of the R\'enyi Entropy.} We first consider the short-time behavior of the R\'enyi entropy dynamics. We apply the short-time expansion to Eq. \ref{wf_time_dependence} and ultilize the relation Eq. \ref{entropy-amplitude}, and to the leading order of entropy change, we obtain
\begin{equation}
\lim\limits_{t\rightarrow 0}\frac{d S^{(2)}}{dt}=2\frac{\langle \Psi_\rho(0)|\hat{H}_\text{d}|\Psi_\rho(0)\rangle}{\langle \Psi_\rho(0)|\Psi_\rho(0)\rangle}. \label{short-time1}
\end{equation}  
The physical meaning of the r.h.s. of Eq. \ref{short-time1} in original system is the fluctuation of the dissipation operators. For instance, if the initial state is a pure state and $\hat{\rho}(0)=|\psi(0)\rangle\langle\psi(0)|$, then $|\Psi_\rho(0)\rangle=|\psi(0)\rangle \otimes|\psi(0)\rangle$, and Eq. \ref{short-time1} can be rewritten as
\begin{align}
&\lim\limits_{t\rightarrow 0}\frac{d S^{(2)}}{dt}=\nonumber \\
&4\sum\limits_{\mu}\gamma_\mu\left(\langle\psi(0)|\hat{L}^\dag_\mu\hat{L}_\mu|\psi(0)\rangle-|\langle \psi(0)|\hat{L}_\mu|\psi(0)\rangle|^2\right). \label{short-time2}
\end{align}
Suppose all $\gamma_\mu$ are taken as the same $\gamma$, this result shows that the time-dependence of $S^{(2)}$ is governed by a dimensionless time $\gamma t$. In other word, the larger $\gamma$ is, the faster the R\'enyi entropy dynamics increases. This $\gamma t$ scaling is shown in Fig. \ref{dynamics_general} for two different models, where one can see that the short-time parts of $S^{(2)}$ curves with different $\gamma$ collapse into a single line when plotted in term of $\gamma t$. The dashed lines compare the short-time behavior with the slope given by Eq. \ref{short-time1} and Eq. \ref{short-time2}. 

In Fig. \ref{dynamics_general}, one also finds that $S^{(2)}$ no longer obeys the $\gamma t$ scaling when $\gamma t>1$. Moreover, in the strong dissipation regime, the insets plotted in term of $tJ$ show an opposite trend at long-time, that is, the larger $\gamma$ is, the slower the R\'enyi entropy increases. In fact, the long-time behavior of $S^{(2)}$ exhibits a $t/\gamma$ scaling. As shown in the insets of Fig. 2, when the long-time part of $S^{(2)}$ curves with different $\gamma$ are ploted in term of $tJ^2/\gamma$, they all collapse into a single curve.   

\textit{Lindblad Spectrum with Strong Dissipation.} This opposite behavior between short- and long-time can be understood very well in term of the Lindblad spectrum. As one can see from Fig. \ref{spectrum}(a,b), for strong dissipation, the main feature of the Lindblad spectrum is that it separates into segments along the imaginary axes of the spectrum, and the separation between segments are approximately $2\gamma$. For each segment, the width along the imaginary axes is approximately given by $J^2/\gamma$. This feature can be understand by perturbation treatment of $\hat{H}_\text{s}-i \hat{H}_\text{d}$. Since the dissipation strength is stronger than the typical energy scales of the Hamiltonian, we can treat $\hat{H}_\text{s}$ as a perturbation to $\hat{H}_\text{d}$. To the zeroth order of $\hat{H}_\text{d}$, the spectrum is purely imaginary and different segments are separated by $2\gamma$. More importantly, it worth emphasizing that the eigenstates of $\hat{H}_\text{d}$ are usually highly degenerate, for instance, when different $\hat{L}_\mu$ commute with each other and are related by a symmetry, such as $\hat{L}_\mu$ being $\hat{n}_i$ in our examples. Usually, $\hat{H}_\text{s}$ and $\hat{H}_\text{d}$ do not commute with each other, and the perturbation in $\hat{H}_\text{s}$ lifts the degeneracy of the imaginary parts and gives rise to a broadening of the order of $J^2/\gamma$, due to the nature of the second order perturbation. 

We call these eigenstates with imaginary energies of the order of a few times of $\gamma$ as ``high imaginary energy states", and these eigenstates with imaginary energies of the order of a few times of $J^2/\gamma$ as ``low-lying imaginary energy states". For a generic initial state, both two types of eigenstates are occupied. Quite generally, the occupations of the ``high imaginary energy states" are significant, for instance, when the initial state is taken as the eigenstates of $\hat{H}_\text{s}$. With the relation between the R\'enyi entropy dynamics and the Lindblad spectrum discussed above, it is clear that the short-time dynamics is dominated by these ``high imaginary energy states" that gives a dynamics scaled by $t\gamma$. Nevertheless, when $\gamma t >1$, the weights on these ``high imaginary energy states" mostly decay out and the long-time dynamics is therefore dominated by the ``low-lying imaginary energy states" that gives a dynamics scaled by $t J^2/\gamma$. 

\begin{figure}[t] 
    \centering
    \includegraphics[width=0.48\textwidth]{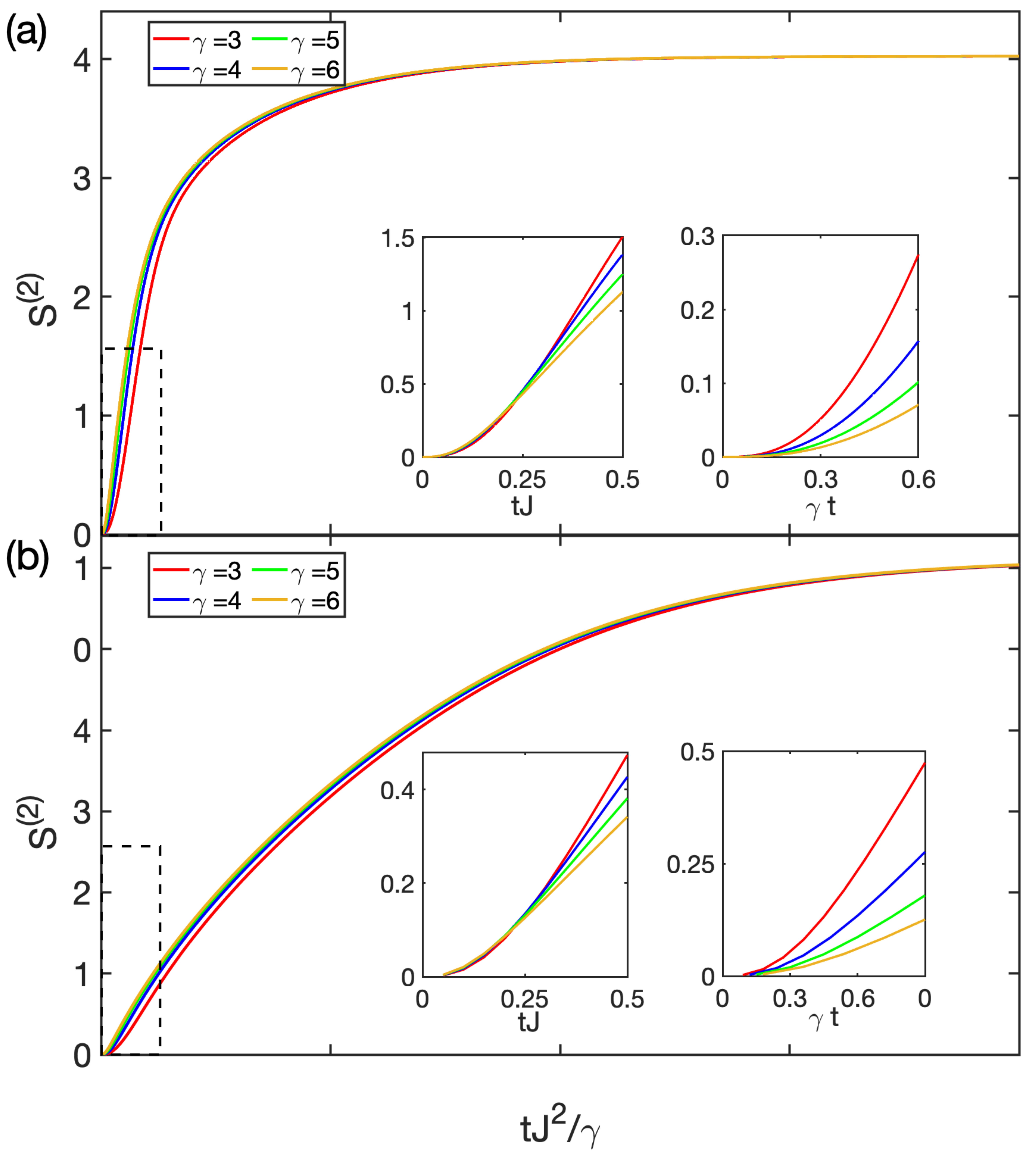}
    \caption{The dynamics of the second R\'enyi entropy $S^{(2)}$ as a function of $t J^2/\gamma$ for specific initial state. $\gamma$ is the dissipation strength. Different curves have different $\gamma$ in unit of $J$. The inset show the short-time behavior of $S^{(2)}$ as functions of $tJ$ and $t \gamma$. (a) is for the Bose-Hubbard model with $U=J$ and the number of sites $L=6$, and the number of bosons $N=3$. (b) is for hard core bosons model with $V=J$, $L=8$ and $N=4$. The initial states are taken as the zero-energy eigenstate of $\hat{H}_\text{d}$, that are $|111000\rangle$ for (a) and $|11110000\rangle$ for (b) in Fock bases. }
     \label{dynamics_specific}
\end{figure}

\textit{Quantum Zeno Effect Revisited.} Here we consider a specific initial state that satisfies $\hat{H}_\text{d}|\Psi(0)\rangle=0$. In other word, such initial states do not exhibit fluctuation of dissipation operators. Thus, according to Eq. \ref{short-time1} and Eq. \ref{short-time2}, the initial slop of $S^{(2)}$ is zero. Moreover, in the strong dissipation regime, the populations of the ``high imaginary energy states" are strongly suppressed by the ``gap" between different segments and their contribution becomes negligible, and such initial states mainly populate the ``low-lying imaginary energy states", as we shown in Fig. \ref{spectrum}(b). Therefore, the entire dynamics of the second R\'enyi entropy is set by the energy scale $J^2/\gamma$ and it obeys the $t/\gamma$ scaling. This is shown in Fig. \ref{dynamics_specific} for two models. To contrast such specific initial states with generic states discussed above, we plot in the inset of Fig. \ref{dynamics_specific} the short-time behavior of $S^{(2)}$ as a function of $t\gamma$ and $tJ$. Unlike the results shown in Fig. \ref{dynamics_general}, the short-time dynamics with $t\gamma<1$ are quite different, because it does not exhibit linear behavior and different curves do not collapse into a single line in term of $t\gamma$. 

For these initial states, that the dynamics is slower with stronger dissipation is reminiscent of the quantum Zeno effect. In fact, the quantum Zeno effect can indeed be understood in this way. Introducing $\{|M\rangle\},(M=1,\dots,\mathcal{D}_\text{H})$ as a set of complete and orthogonal measurement bases, we define the projection operators as $\hat{P}_M=|M\rangle\langle M|$, and the frequent measurement process can also be described by the Lindblad equation Eq. \ref{Lindblad} with dissipation operator $\hat{L}_\mu$ given by all $\hat{P}_M$. With such dissipation operators, the Lindblad spectrum exhibits a set of ``low-lying imaginary energy states" with energy scale given by $J^2/\gamma$. It can be shown that, as long as the initial state density matrix is diagonal in the measurement bases, the initial states satisfy $\hat{H}_\text{d}|\Psi(0)\rangle=0$.  

\textit{From Strong to Weak Dissipation.} Finally we show that when $\gamma$ decreases and eventually becomes weaker compared with the typical energy scales in the Hamiltonian, the segments structure in the Lindblad spectrum disappears, as we shown in Fig. \ref{spectrum}(c). Thus, the entropy dynamics for generic states no longer display the feature of two time scales. The quantum Zeno effect also disappears even for the specific initial states, and this is understandable because in this regime, the typical time interval between two measurements is already longer than the intrinsic evolution time of the system. 

\textit{Summary.} In this work, we establish a relation between the R\'enyi entropy dynamics and the Lindblad spectrum in double space. At the strong dissipation regime, the Lindblad spectrum exhibits a segment structure, in which we can introduce the ``high imaginary energy eigenstates" and the ``low-lying imaginary energy eigenstates". For a generic initial state with significantly occupied ``high imaginary energy eigenstates", the former dominates the short-time dynamics and the latter dominates the long-time dynamics, which respectively give rise to $t\gamma$ scaling and $t/\gamma$ scaling. For a specific initial state with only ``low-lying imaginary energy eigenstates" significantly occupied, the dynamics is dominated by $t/\gamma$ scaling, and we show the quantum Zeno effect belongs to this class. We illustrate our results with two concrete models. The second R\'enyi entropy can now been measured in ultracold atomic gases in optical lattices, and in fact, it has been measured in the Bose-Hubbard model with or without disorder \cite{Greiner1,Greiner2,Greiner3}. The dissipation operators and their strenghes can also now be controlled in ultracold atomic gases \cite{BHMExp}, our predictions can therefore be verified directly in the experimental setup.    

\textit{Acknowledgment.} We thank Lei Pan, Tian-Shu Deng, Tian-Gang Zhou and Pengfei Zhang for helpful discussions. This work is supported by Beijing Outstanding Young Scientist Program, NSFC Grant No. 11734010, MOST under Grant No. 2016YFA0301600.

\textit{Note Added.} When finishing this work, we become aware of a work in which similar behaviors of the Lindblad spectrum in strong dissipation regime are also discussed \cite{full_spectrum}.


\begin{thebibliography}{99}

\bibitem{open}
H. P. Breuer and F. Petruccione, The Theory of Open Quantum Systems (Oxford University Press, Oxford, 2007).

\bibitem{Pan}
L. Pan, X. Chen, Y. Chen, and H. Zhai, Nat. Phys. {\bf 16}, 767(2020).


\bibitem{Zeno_paradox}
B. Misra and E. C. G. Sudarshan, Journal of Mathematical Physics {\bf 18}, 756 (1977).

\bibitem{Zeno1990}
W. M. Itano, D. J. Heinzen, J. J. Bollinger, and D. J. Wineland, Phys. Rev. A {\bf 41}, 2295 (1990).

\bibitem{quantum_Zeno}
A. G. Kofman and G. Kurizki, Nature {\bf 405}, 546 (2000).

\bibitem{Zeno_review}
K. Koshino and A. Shimizu, Phys. Rep. {\bf 412}, 191 (2005).

\bibitem{entang_review}
L. Aolita, F. de Melo and L. Davidovich, Rep. Prog. Phys. {\bf 78}, 042001 (2015)

\bibitem{Qi_Cricuit}

P. Lorenzo, S. Christoph, and X.-L., Qi, J. High Energ. Phys. \textbf{2020}, 63 (2020).

\bibitem{Chen}
Y.~Chen, X.-L. Qi and P.~Zhang, J. High Energ. Phys. \textbf{2020}, 121 (2020).

\bibitem{Kitaev}
P. Dadras, A. Kitaev, arXiv: 2011.09622

\bibitem{YuChen}
Y. Chen, arXiv: 2012.00223

\bibitem{Zhai}
K. Su, P. Zhang and H. Zhai, arXiv: 2101.*****

\bibitem{Operator_Schmidt}
J. E. Tyson, J. Phys. A: Math. Gen. {\bf 36}, 10101 (2003).

\bibitem{Mixed_state}
M. Zwolak and G. Vidal, Phys. Rev. Lett. {\bf 93}, 207205 (2004).

\bibitem{Prosen1}
T. Prosen, Phys. Rev. Lett. {\bf 109}, 090404 (2012)

\bibitem{Prosen2}
M. V. Medvedyeva, F. H. L. Essler, T. Prosen, Phys. Rev. Lett. {\bf 117}, 137202 (2016)

\bibitem{Universal_spectra}
S. Denisov, T. Laptyeva, W. Tarnowski, D. Chru?ci?ski, and K. Zyczkowski, Phys. Rev. Lett. {\bf 123}, 140403 (2019)


\bibitem{tenfold}
S. Lieu, M. McGinley, and N. R. Cooper, Phys. Rev. Lett. {\bf 124}, 040401 (2020).


\bibitem{local_random_Liouvillians}
 K. Wang, F. Piazza, and D. J. Luitz, Phys. Rev. Lett. {\bf 124}, 100604 (2020).

\bibitem{Wang}
D. Yuan, H. Wang, Z. Wang, D. L. Deng, arXiv: 2009.00019

\bibitem{Greiner1}
R. Islam, R. Ma, P. M. Preiss, M. Eric Tai, A. Lukin, M. Rispoli, M. Greiner, Nature {\bf 528}, 77 (2015).

\bibitem{Greiner2}
A. M. Kaufman, M. Eric Tai, A. Lukin, M. Rispoli, R. Schittko, P. M. Preiss, M. Greiner, Science {\bf 353}, 794 (2016)

\bibitem{Greiner3}
A. Lukin, M. Rispoli, R. Schittko, M. Eric Tai, A. M. Kaufman, S. Choi, V. Khemani, J. L\'eonard, M. Greiner, Science {\bf 364}, 256 (2019)



\bibitem{BHMExp}
R. Bouganne, M. B. Aguilera, A. Ghermaoui, J. Beugnon, F. Gerbier, Nat. Phys. {\bf 16}, 2125 (2020).

\bibitem{full_spectrum}
V. Popkov and C. Presilla, arXiv:2101.05708 


\end{thebibliography}
\end{document}